\documentclass[journal,comsoc]{IEEEtran}
\usepackage[T1]{fontenc}
\usepackage{color}
\usepackage{graphicx}
\usepackage[caption=false,font=footnotesize]{subfig}
\usepackage{fixltx2e}

\ifCLASSINFOpdf
\else
\fi
\usepackage{amsmath}
\usepackage[cmintegrals]{newtxmath}
\begin{document}
%
\title{Control of Orbital Angular Momentum Regimen by Modulated Metasurface Leaky-Wave Antennas}
%
%
%

\author{Amrollah~Amini,
        and~Homayoon~Oraizi,~\IEEEmembership{Life~Senior,~IEEE,}
\thanks{A. Amini and H. Oraizi are with the School
of Electrical Engineering, Iran University of Science and Technology, Tehran 1684613114,
Iran, e-mail: (h\_oraizi@iust.ac.ir).}}

%
%

\markboth{}%
{Shell \MakeLowercase{\textit{et al.}}: Bare Demo of IEEEtran.cls for IEEE Communications Society Journals}
%



\maketitle

\begin{abstract}
In this paper, we present the design procedure of  modulated metasurface leaky-wave antennas (MMLWAs) to generate  coaxial superposition of vortex beams with several orbital angular momentum (OAM) states.
Based on the Flat Optics (FO) technique and aperture field estimation (AFE) method, an analytical framework is proposed to facilitate the implementation of MMLWAs generating multiple topological charges in the OAM regimen. 
Furthermore, using the spectral analysis which has been derived from the proposed model, we have shown that the symmetry of aperture shape can affect the purity of the mode. Also the perfectly symmetric circular shape is introduced as  an ideal choice for high-purity vortex generation.
This single aperture antenna with embedded monopole feed can be an appropriate alternative to more complex vortex beam generators such as spiral phase plates and circular antenna arrays.
\end{abstract}


\IEEEpeerreviewmaketitle

\section{Introduction}
\IEEEPARstart{I}{n} addition to energy and spin angular momentum (SAMs), the electromagnetic waves can also carry different orbital angular momentum (OAM) modes.
The practical use of beams carrying OAM eigenstates was first proposed by Allen et al in the optical regime \cite{allen_1992}.
Owing to the orthogonality of OAM states, recently, the use of this fascinating feature of electromagnetic waves at radio frequencies has received much attention \cite{thide_2007}, which  can be utilized to enhance the channel capacity of wireless communication networks \cite{chen_2020}.
During the past decades, many attempts have been made to generate electromagnetic waves carrying OAM modes. Spiral phase plates (SPPs) \cite{hui_2015,chen_2016}, antenna arrays with circular arrangements \cite{gao_2014}, all-dielectric metamaterials \cite{yi_2019}, and metasurfaces  \cite{yu_2016}-\cite{karimipour_2019} are among the most common structures to realize OAM wave generators.
To further increase the channel capacity, the use of multi-mode generators has recently been suggested. 
Based on this idea, several methods have been reported to control the multi-mode OAM waves in the topological charge space, such as  utilizing shape-tailored metasurfaces \cite{yang_2021}, \cite{yang_2021_tap},  quadratic phase plates (QPPs) \cite{liu_2018}, pinhole plates \cite{yang_2019_pinhole}, plasmonic metasurfaces \cite{zhang_2019} and planar spiral phase plates (SPPs) \cite{cheng_2014}. The structures proposed in \cite{yang_2021}-\cite{cheng_2014} are based on transmissive/reflective metasurfaces. 
Despite the advantages of transmissive/reflective metasurfaces, they are limited by their need for protruded feed networks for illumination, which can make them difficult to integrate into planar devices.
An alternative approach to address this challenge is to exploit metasurface-based leaky-wave antennas with embedded feed systems.

Modulated metasurface leaky-wave antennas (MMLWAs) have recently attracted considerable attention, owing to their high efficiency, low profile structures, and simple manufacturing process. Since their aperture distribution can be readily controlled, they are able to synthesize beams with different shapes and spin angular momentum states \cite{minatti_2012}-\cite{caminita_2021}.
In addition to linear momentum and SAM modes,  excitation of orbital angular momentum  modes using MMLWAs have also been considered in recent years \cite{meng_2019}-\cite{amini_2021}. 
The advantage of metasurface-based leaky-wave OAM radiators over their transmissive \cite{zhang_2018, jiang_2018}  and reflective \cite{ding_2019, yang_2020}  counterparts is their surface-mounted feeding networks, which significantly reduce the overall size of such structures and make them suitable for use in planar systems. Furthermore, thanks to the controllability of the leakage factor of MMLWAs, there is an additional degree of freedom in improving the beam convergence. 

Despite the above-mentioned advantages, the full-wave analysis and design of MMLWAs may be a challenging task due to the use of high density constituent sub-wavelength unit cells. The dimension of the unit cell is typically selected in the range of $\lambda/10$ to $\lambda/5$ ($\lambda$ as the free-space wavelength), whereas the overall dimension of antenna may be $10\lambda$ to $20\lambda$. That is, in order to distinguish each inclusion, the full-wave simulator should apply dense meshing, which drastically increases  the analysis time.
To avoid the full-wave simulation, several methods have been proposed for the synthesis and design of  MMLWAs. Holographic technique \cite{fong_2010}, aperture field estimation method \cite{minatti_2015}-\cite{amini_2019}, Flat Optics (FO) \cite{minatti_FO_2016}, \cite{minatti_2016}, and the Method of Moments (MoM) \cite{ovejero_2015}-\cite{bodehou_2019}, are among the well-established analytical models.
All the above methods use the homogenized impedance boundary condition for the analysis of MMLWAs. Thus, the quasi-periodic array of constituent pixels can be viewed as the equivalent surface impedance tensor. Using this paradigm significantly reduces the complexity of the analysis, which can be applied for large scale MMLWAs.

In this paper, the combination of Flat Optics and aperture field estimation technique is exploited to analyze MMLWAs generating multiple modes in the OAM regimen. Based on this method, we can provide a simple and straightforward solution for predicting the mode distribution of the vortex wave radiated by MMLWAs. 
Due to the superiority of leaky-wave structures over the transmitarrays and reflectarrays, in this paper, the process of analyzing and designing MMLWAs is proposed to control the OAM spectrum.
\section{Flat optics and aperture field estimation technique}
\begin{figure}
\centering
\includegraphics[width = 0.37\textwidth]{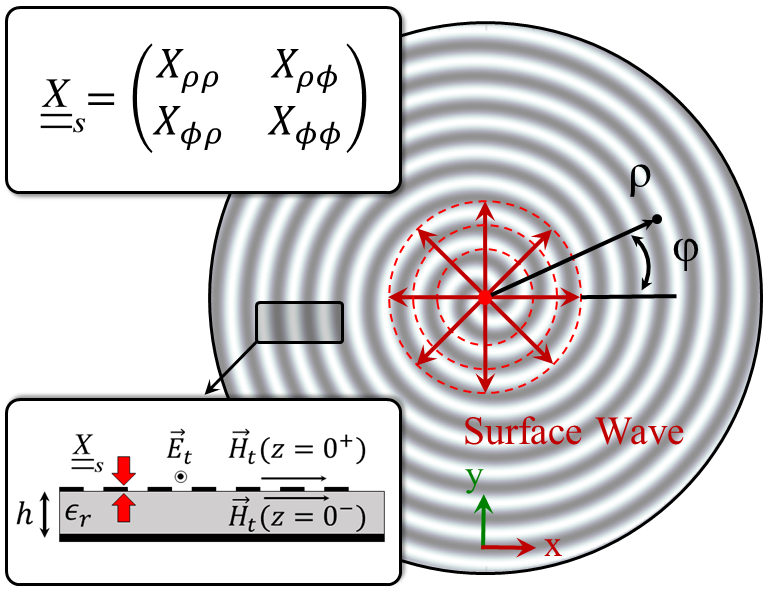}
\caption{Geometry of a  modulated metasurface leaky-wave  antenna consisting of a vertical monopole located at the origin. The red arrows in the left inset indicate the transparent surface impedance, which excludes the grounded dielectric substrate.}
\label{fig:Fig1}
\end{figure}

The Flat Optics method based on the adiabatic Floquet-wave expansion is an effective method for the analysis of MMLWAs, which was originally proposed by Minatti et al \cite{minatti_FO_2016}. Owing to its high accuracy and speed, FO can be widely used in the design of large-scale MMLWAs for generating beams with arbitrary spin \cite{minatti_2016} and orbital angular momentums \cite{amini_2021}.
A schematic representation of a modulated metasurface antenna is shown in Fig.\ref{fig:Fig1}, which is fed by a vertical monopole antenna embedded in the metasurface plane. 
The metasurface can be characterized by the equivalent surface impedance boundary condition (as shown in Fig.\ref{fig:Fig1}, left inset). By applying an appropriate modulation to the surface reactance, the conversion of the surface wave into the leaky wave will occur effectively.
In the most general case, the tensorial surface reactance required to support an aperture electric field (called $\vec{E}_a(\vec{\rho}) = E_{ax}(\vec{\rho}) \hat{x} + E_{ay}(\vec{\rho}) \hat{y}$) can satisfies the following relation \cite{minatti_2015}:
\begin{equation}
\underline{\underline{X}}_s(\vec{\rho})\cdot \hat{\rho} = X_\rho[\hat{\rho} + 2 Im\{\frac{\vec{E}_{a}(\vec{\rho})}{-E_0H_1^{(2)}(\tilde{k}_{sw}\rho)}\}]
\label{eq:Xs_Eap}
\end{equation}
where $H_1^{(2)}(\tilde{k}_{sw}\rho)$ is the Hankel function of the second kind and denotes the surface wave excited by the vertical monopole. The complex surface wave number $\tilde{k}_{sw}=\tilde{\beta}_{sw} - j\tilde{\alpha}_{sw}$ is unknown, and its value will be calculated using FO method.
Equation (\ref{eq:Xs_Eap}), determines the modulation parameters  according to the desired aperture field, leading to far-zone field with the desired SAM and OAM states. For example in order to have far-zone phase distribution with superimposed OAM states in the $\theta = \theta_0$ and $\phi = \phi_0$ spherical angles, the aperture field components may be defined as follows:
\begin{equation}
\begin{split}
E_{ax, ay}(\vec{\rho}) = M_{x, y}(\vec{\rho}) \frac{E_0}{\sqrt{2\pi\rho|\tilde{k}_{sw}|}}e^{-\tilde{\alpha}_{sw}\rho} e^{-jk\rho\sin \theta_0 \cos (\phi - \phi_0) - j\zeta_0}
\end{split}
\label{eq_Eax_ay}
\end{equation}
where $M_x(\vec{\rho})$ and $M_y(\vec{\rho})$ are defined as 
\begin{equation}
M_{x,y}(\vec{\rho}) = M_{0x, 0y}\sum_{l=L_1}^{L_2}c_l e^{-jl\phi}
\end{equation}
Note that  $l$ can be interpreted as $l$'th harmonic in the OAM spectrum space.
In the above equation $M_{0x}$ and $M_{0y}$ are the degrees of freedom and can have arbitrary values. However, the spin angular momentum of the far-zone field imposes some restrictions on their values, which their appropriate selections  will be discussed in the next section. Substituting (\ref{eq_Eax_ay}) in (\ref{eq:Xs_Eap}) yields
\begin{equation}
  \underline{\underline{X}}_s(\vec{\rho}) = X_{\rho\rho}(\vec{\rho})\hat{\rho}
\hat{\rho} + X_{\rho\phi}(\vec{\rho})(\hat{\rho}\hat{\phi} + \hat{\phi}\hat{\rho}) + X_{\phi\phi}(\vec{\rho})\hat{\phi}\hat{\phi}
\label{eq:Xs}
\end{equation}
\begin{equation}
X_{\rho\rho}(\vec{\rho}) = X_\rho [1+m_\rho(\vec{\rho}) \cos (Ks(\vec{\rho}) + \Phi_\rho(\vec{\rho}))]
\label{eq:X_rho_rho}
\end{equation}
\begin{equation}
X_{\rho\phi}(\vec{\rho}) = X_\rho m_\phi(\vec{\rho}) \cos(Ks(\vec{\rho}) + \Phi_{\phi}(\vec{\rho}))
\label{eq:X_rho_phi}
\end{equation}
where:
\begin{equation}
Ks(\vec{\rho}) = \tilde{\beta}_{sw}\rho - k\rho \sin \theta_0 \cos(\phi - \phi_0)
\label{eq:Ks}
\end{equation}
\begin{equation}
\Phi_\rho(\vec{\rho}) = -arg\{M_x(\vec{\rho})\ \cos \phi + M_y(\vec{\rho}) \sin \phi\}
\end{equation}
\begin{equation}
\Phi_\phi(\vec{\rho}) = -arg\{-M_x(\vec{\rho})\ \sin \phi + M_y(\vec{\rho}) \cos \phi\}
\end{equation}
\begin{equation}
m_\rho(\vec{\rho}) = |M_x(\vec{\rho}) \cos \phi + M_y(\vec{\rho}) \sin \phi|
\end{equation}
\begin{equation}
m_\phi(\vec{\rho}) = |-M_x(\vec{\rho}) \sin \phi + M_y(\vec{\rho}) \cos \phi|
\end{equation}
Observe that the modulation of $X_{\phi\phi}$ in (\ref{eq:Xs}) is not synthesized explicitly in the formulations, due to the design scheme introduced in \cite{minatti_2015} which is based on the quasi-TM field synthesis. However, in order to Floquet-wave expansion can be applicable, we define $X_{\phi\phi}$ as follows \cite{minatti_FO_2016}:
\begin{equation}
X_{\phi\phi}(\vec{\rho}) = X_\rho [1-m_\rho(\vec{\rho}) \cos (Ks(\vec{\rho}) + \Phi_\rho(\vec{\rho}))]
\label{eq:X_phi_phi}
\end{equation}
Note that, $Ks(\vec{\rho})$ is the rapidly varying part and $\Phi_{\rho, \phi}(\vec{\rho})$ are the slowly varying parts of the modulation phase, satisfying the following inequality:
\begin{equation}
|\nabla_{\vec{\rho}}Ks(\vec{\rho})| \gg |\nabla_{\vec{\rho}}\Phi_{\rho, \phi}(\vec{\rho})|
\end{equation}
In (\ref{eq:X_rho_rho}) and (\ref{eq:X_rho_phi}), $m_\rho(\vec{\rho})$ and $m_\phi(\vec{\rho})$ indicate the modulation depth, and $X_\rho$ and $X_\phi$ are the average reactances.
It is worthy to note that, the modulation parameters are independent of $\tilde{\alpha}_{sw}$, but their dependence on the propagation constant ($\tilde{\beta}_{sw}$) can be seen in Equation (\ref{eq:Ks}). The propagation constant can be approximated by transverse resonance method  \cite{martini_2015}, which is accurate enough for our synthesis purpose. Thus we have:
\begin{equation}
  \tilde{\beta}_{sw} \approx k \sqrt{1 + (\frac{X_0}{\eta_0})^2}
\end{equation}
where $X_0$ is calculated by the following nonlinear equation:
\begin{equation}
  X_0 = X_\rho [1 - \frac{X_0 \epsilon_r \cot (kh \sqrt{\epsilon_r - 1 - (X_0/\eta_0)^2})}{\eta_0 \sqrt{\epsilon_r - 1 - (X_0/\eta_0)^2}}]
\end{equation}
\subsection{Calculating $\tilde{k}_{sw}$ using the Flat Optics method}
After determining the modulation parameters, the complex wavenumber ($\tilde{k}_{sw}$) can be calculated through Flat Optics analytical framework. In the presence of modulation, the surface current must be expanded in terms of higher order Floquet modes \cite{minatti_FO_2016}, thus, we have 
\begin{equation}
  \vec{J} = \sum_n \vec{J}^{(n)} = \sum_n \vec{j}^{(0)}e^{-jnKs(\vec{\rho})}H_1^{(2)}(\tilde{k}_{sw}\rho)
  \label{eq:J}
\end{equation}
Using asymptotic form of Hankel function and obtaining derivation of the phase of n-indexed term in Eq. (\ref{eq:J}), we can extract the spatial frequency of the Floquet mode:
\begin{equation}
  \vec{k}^{(n)} = \nabla_{\vec{\rho}}[\tilde{k}_{sw}\rho + n Ks(\vec{\rho})]
\end{equation}
For 0-index mode we have
\begin{equation}
  \vec{k}^{(0)} = \frac{\partial}{\partial \rho} (\tilde{k}_{sw} \rho) \hat{\rho} + \frac{1}{\rho} \frac{\partial}{\partial \phi}(\tilde{k}_{sw}\rho)\hat{\phi}
  \label{eq:k0_vec}
\end{equation}
In (\ref{eq:k0_vec}), it can implicitly be assumed that the 0-index Floquet mode propagates only along $\hat{\rho}$ \cite{minatti_FO_2016} (has cylindrical wavefront), thus we can write:
\begin{equation}
  \vec{k}^{(0)} \approx k^{(0)}\hat{\rho} =  \frac{\partial}{\partial \rho} (\tilde{k}_{sw} \rho) \hat{\rho}
\end{equation}
The impedance boundary condition imposes the following relation:
\begin{equation}
  \vec{E}_t(\vec{\rho}) = \sum_n \vec{E}^{(n)} = \sum_n j \underline{\underline{X}}_s(\vec{\rho}) \cdot \vec{J}^{(n)}
  \label{eq:Et}
\end{equation}
Substuiting (\ref{eq:X_rho_rho}), (\ref{eq:X_rho_phi}) and (\ref{eq:X_phi_phi}) in (\ref{eq:Et}) and applying some mathematical simplifications, yields
\begin{equation}
  \vec{E}^{(n)} = j[\underline{\underline{X}}^{(0)} \cdot \vec{J}^{(n)} + \underline{\underline{X}}^{(-1)} \cdot \vec{J}^{(n + 1)} + \underline{\underline{X}}^{(+1)} \cdot \vec{J}^{(n - 1)}]
  \label{eq:En}
\end{equation} 
where
\begin{equation}
  \underline{\underline{X}}^{(0)} = X_\rho \hat{\rho}\hat{\rho} + X_\phi \hat{\phi}\hat{\phi}
\end{equation}
\begin{equation}
  \begin{split}
  \underline{\underline{X}}^{(\pm 1)} = \frac{1}{2} e^{\mp jKs(\vec{\rho})}[m_\rho(\vec{\rho})e^{\mp j \Phi_\rho(\vec{\rho})}(X_\rho \hat{\rho}\hat{\rho} - X_\phi \hat{\phi} \hat{\phi})   +   \\
  m_\phi(\vec{\rho})e^{\mp j \Phi_\phi(\vec{\rho})} X_\rho (\hat{\rho}\hat{\phi} + \hat{\phi}\hat{\rho})]
  \end{split}
\end{equation}
On the other hand, the spectral Green's function of the grounded dielectric slab establishes the following relation between the electric field and surface current \cite{minatti_FO_2016}:
\begin{equation}
  \vec{E}^{(n)} = -j[\underline{\underline{X}}_0^{-1}(\beta^{(n)}) + \underline{\underline{X}}_g^{-1}(\beta^{(n)})]^{-1} \cdot \vec{J}^{(n)}
  \label{eq:En_GF}
\end{equation}
where
\begin{equation}
    \underline{\underline{X}}_0(\beta^{(n)}) = -\eta_0\frac{\sqrt{(\beta^{(n)})^2 - k^2}}{k}\hat{\rho}\hat{\rho} + 
    \eta_0 \frac{k}{\sqrt{(\beta^{(n)})^2 - k^2}} \hat{\phi}\hat{\phi}
\end{equation}
\begin{equation}
  \begin{split}
    \underline{\underline{X}}_g(\beta^{(n)}) = \eta_0[\frac{\sqrt{\epsilon_rk^2 - (\beta^{(n)})^2}}{\epsilon_r k} \hat{\rho}\hat{\rho} + \frac{k}{\sqrt{\epsilon_r k^2 - (\beta^{(n)})^2}} \hat{\phi}\hat{\phi}] \times \\
    \tan(h\sqrt{\epsilon_r k^2 - (\beta^{(n)})^2})
  \end{split}
\end{equation}
Using (\ref{eq:En}) and (\ref{eq:En_GF}) we can conclude that
\begin{equation}
  \begin{split}
    ([\underline{\underline{X}}_0^{-1}(\beta^{(n)}) + \underline{\underline{X}}_g^{-1}(\beta^{(n)})]^{-1} + \underline{\underline{X}}^{(0)})\cdot\vec{J}^{(n)} + \\
    \underline{\underline{X}}^{(-1)}\cdot \vec{J}^{(n+1)} + \underline{\underline{X}}^{(+1)}\cdot \vec{J}^{(n-1)} = 0
  \end{split}
  \label{eq:eigen_mode_equation}
\end{equation}
To solve this recursive equation we should truncate the number of propagating Floquet modes. For leaky-wave structures, it is sufficiently accurate to consider 3 modes (namely $n = 0, \pm1$). Applying some mathematical simplifications, we can rewrite:
\begin{equation}
  (\underline{\underline{X}}_s^{(0)} - \underline{\underline{X}}^{(-1)}\cdot [\underline{\underline{X}}_s^{(+1)}]^{-1} \cdot \underline{\underline{X}}^{(+1)}  - \underline{\underline{X}}^{(+1)}\cdot [\underline{\underline{X}}_s^{(-1)}]^{-1} \cdot \underline{\underline{X}}^{(-1)}) \cdot \vec{J}^{(0)} = 0
  \label{eq:Eigen_mode}
\end{equation}
where
\begin{equation}
  \underline{\underline{X}}_s^{(n)} = [\underline{\underline{X}}_0^{-1}(\beta^{(n)}) + \underline{\underline{X}}_g^{-1}(\beta^{(n)})]^{-1} + \underline{\underline{X}}^{(0)}
\end{equation}
Equation (\ref{eq:Eigen_mode}) is an eigenmode equation that by solving it and obtaining its eigenvalues, the surface wave number can be exactly calculated.
Using (\ref{eq:En_GF}) and (\ref{eq:eigen_mode_equation}) the leaky-mode field can also be obtained using the following equation:
\begin{equation}
  \vec{E}^{(-1)} = j [\underline{\underline{X}}^{(-1)} - \underline{\underline{X}}^{(0)} \cdot (\underline{\underline{X}}_s^{(-1)})^{-1} \cdot \underline{\underline{X}}^{(-1)}] \cdot \vec{J}^{(0)}
\end{equation}
After determining the aperture field ($\vec{E}^{(-1)}$) we can calculate the far-zone field using Fourier transformation \cite{amini_2021}. 
\subsection{Calculation of mode purity}
In order to quantitatively study the mode purity of the radiated vortex beam, we need to analyze the characteristics of the beam  in the topological charge space using Fourier transformation. The corresponding equation for mode purity can be defined as \cite{jack_2008}:
\begin{equation}
  A_l = \frac{1}{2\pi} \int_0^{2\pi} \psi_{co} (\theta_{max}, \phi) e^{+jl\phi} d\phi
\end{equation}
where $\psi_{co}(\theta, \phi)$ is the co-polarization component of far-zone field, and the spherical angle $\theta_{max}$ is the inclination from $z$ axis which the co-polarization gain is maximum.
\section{Antenna design}
To investigate the application of the analytical model, In this section OAM wave generators with right-hand polarizations are designed. 
Without loss of generality, the design frequency and antenna dimensions are selected as 18 GHz and $14\lambda\times 14 \lambda$, respectively ($\lambda$ as wavelength in the free space). The radiated beam is directed towards the broadside direction ($\theta_0 = 0^\circ$). 
\subsection{Single-mode generation}
\begin{figure}
  \centering
  \includegraphics[width = 0.5\textwidth]{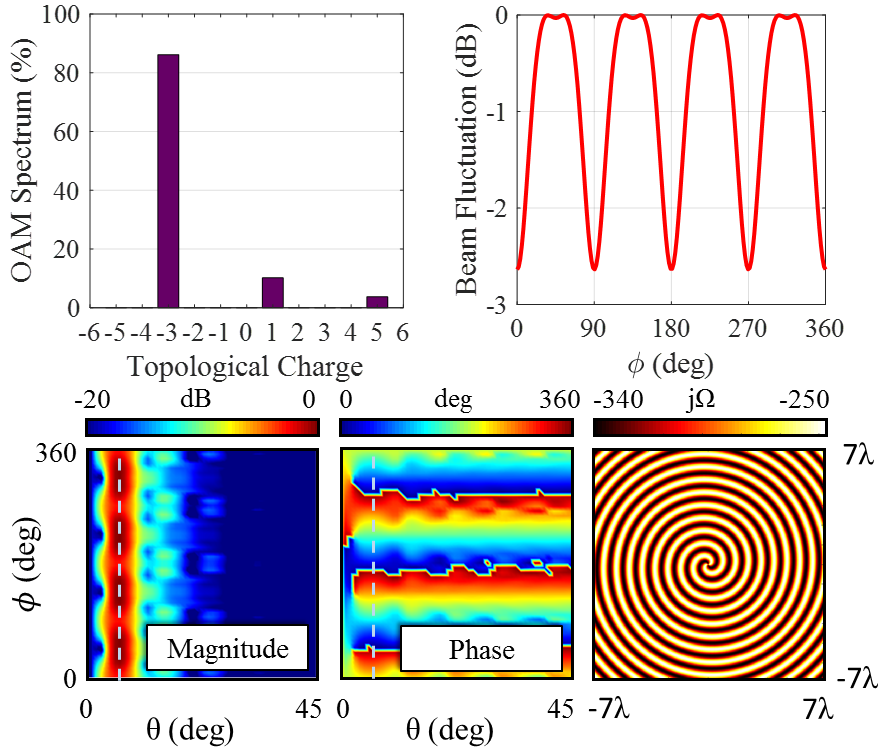}
  \caption{Mode purity and radiation patterns of rectangular shaped MMLWA generating polarized OAM wave with topological charge of -3.  }
  \label{fig:Fig2}
  \end{figure}

\begin{figure}
  \centering
  \includegraphics[width = 0.5\textwidth]{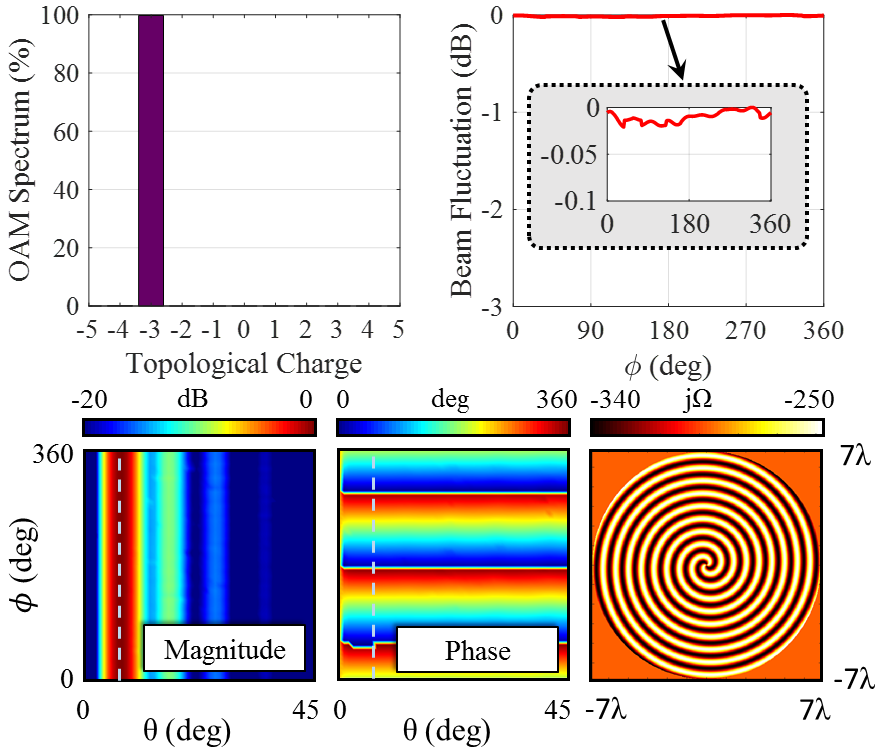}
  \caption{Mode purity and radiation patterns of circular shaped MMLWA generating polarized OAM wave with topological charge of -3.}
  \label{fig:Fig3}
  \end{figure}

To generate pure OAM modes and minimize beam fluctuations, the circular shape is selected as the radiation aperture. In Figs \ref{fig:Fig2} and \ref{fig:Fig3},  the simulation results for the rectangular and circular shapes are shown. Both structures in Figs \ref{fig:Fig2} and \ref{fig:Fig3} are supposed to generate a single mode OAM wave with topological charge of $l = -3$.   Observe that, for the circular shape the mode purity reaches to 100\%, and beam fluctuation is minimum. 
The reason for the higher purity of the circular aperture can be explained from the analytical model as follows.

The relationship between the aperture field and the far-zone field can be expressed by Fourier transform \cite{amini_2021}:
\begin{equation}
f_{x, y} = \iint_{ap} E_{ax, ay}(\rho ', \phi ') e^{jk\rho ' \sin \theta \cos (\phi - \phi ')} \rho ' d\rho ' d \phi '
\label{eq:fx_fy}
\end{equation}
Substituting (\ref{eq_Eax_ay}) in (\ref{eq:fx_fy}) yields:
\begin{equation}
\begin{split}
f_{x, y} = \iint_{ap} M_{0x, 0y} e^{-jl \phi '}\frac{E_0}{\sqrt{2\pi \rho ' |\tilde{k}_{sw}|}} e^{-\tilde{\alpha}_{sw}\rho '} \times \\
e^{jk\rho ' \sin \theta \cos (\phi - \phi ')} e^{-j\zeta_0} \rho ' d \rho ' d\phi '
\end{split}
\label{eq:fx_fy_integration}
\end{equation}
Note that the radiation is considered to be single-mode with topological charge of $l$ at $\theta_0 = 0^\circ$. 
Using the substitution rule in (\ref{eq:fx_fy_integration}) by setting $u = \phi - \phi '$, we conclude that:
\begin{equation}
f_{x, y} = e^{-jl\phi} \iint_{ap} M_{0x, 0y} \frac{e^{jl u}}{\sqrt{2\pi \rho ' |\tilde{k}_{sw}|}} e^{jk\rho ' \sin \theta \cos u} \rho ' d\rho ' du
\label{eq:fx_fy_u}
\end{equation}
For circular apertures, the integration interval corresponding to variable u is  $[\phi, \phi - 2\pi]$. Due to the fact that the integrand is periodic (with period of $2\pi$), the integration interval can be reduced to $[0, -2\pi]$. Observe that in (\ref{eq:fx_fy_u})  the double integration is independent of $\phi$  and $f_x$ and $f_y$ only  have an angular dependency of the form of $exp(-jl\phi)$. 
Accordingly, it is theoretically possible to generate OAM-carrying waves with $100\%$ mode purity by circular apertures.
Note that, due to the angular dependence of double integration for rectangular apertures, we have some degradation of mode purity (as shown in Fig.\ref{fig:Fig2}). This dependence causes higher order modes (with small amplitudes) to appear in the radiation beam.
Note that, for both cases, the phase singularity occurs at $\theta = 0^\circ$.
  \begin{figure*}[h]
    \centering
        \subfloat[\label{fig:Fig_Ex_Plus3}]{
          \includegraphics[width = 0.5\textwidth]{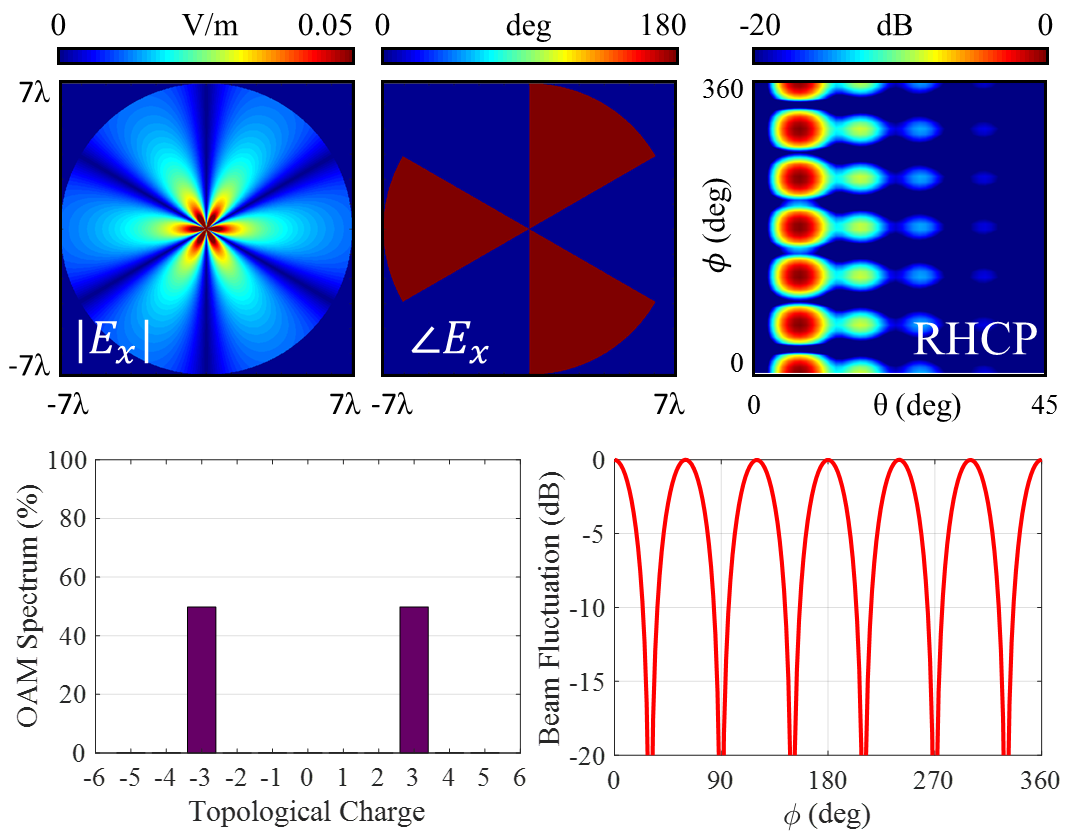}
        }
        \subfloat[\label{fig:Fig_Ex_Plus4}]{
          \includegraphics[width = 0.5\textwidth]{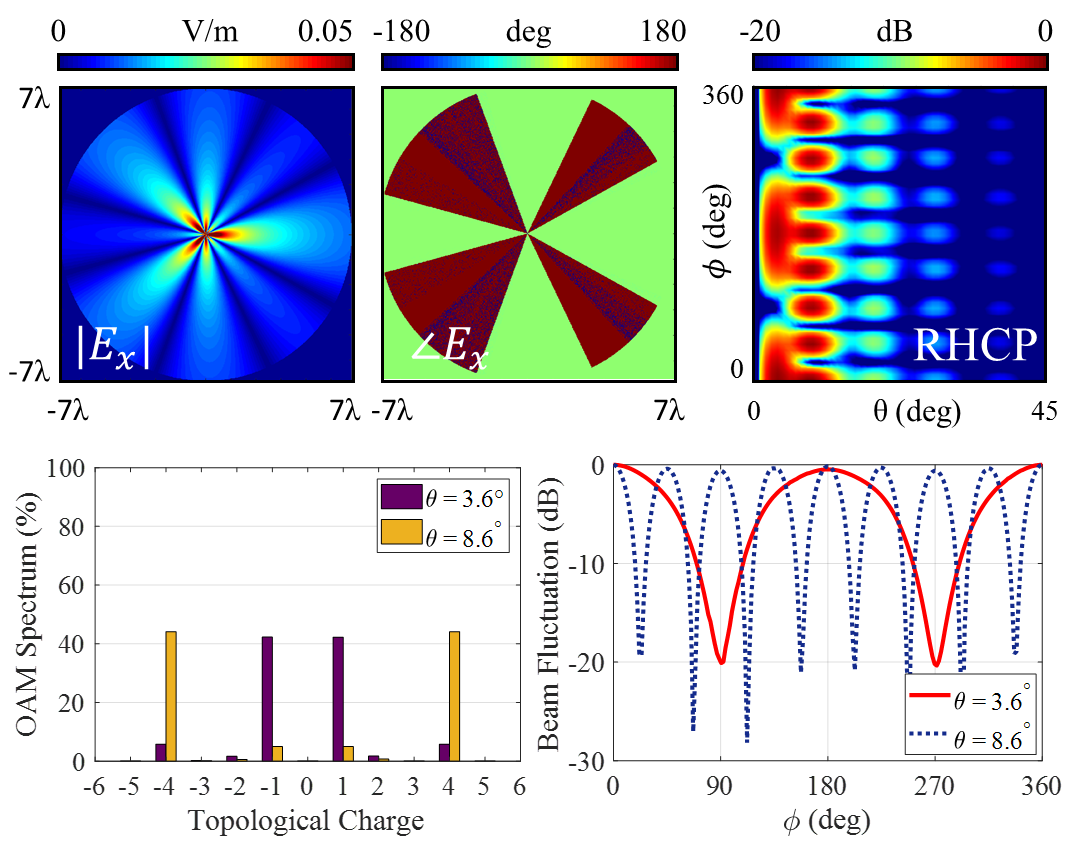}
        }
        \caption{Analytical results for aperture field distribution, far-zone pattern, mode purity, and beam fluctuation of proposed modulated metasurface to generate OAM waves with topological charges of (a) $l = \pm 3$ and (b) $l = \pm 1, \pm 4$. } 
  \end{figure*}
\subsection{\label{subsec:multi_mode}Multi-mode generation}
A simple way to generate superimposed OAM states is to tailor the shape of radiating aperture. In \cite{yang_2021} the envelope-modulated scheme is used to deform the aperture shape and consequently distribute arbitrary OAM modes in the spectrum. The results in Figs \ref{fig:Fig2} and \ref{fig:Fig3} confirm this statement. Observe that, imposing some rotational asymmetries (for example, utilizing shapes other than perfectly symmetrical circular boundaries) distributes power in other modes. However, in this method, selecting the desired modes in the spectrum requires some optimization algorithms, which complicates the design process.
In this paper we manipulate the  distribution of modulation depth to generate multi mode OAM waves with arbitrary topological charges.
For example, in order for power to be distributed in modes $l = 3, l = -3$, the modulation depths must have the following form:
\begin{equation}
  M_{x, y} = M_{0x, 0y} (c_3 e^{-j3\phi} + c_{-3} e^{j3\phi})
\end{equation} 
Given that the aperture fields are multiplied by these modulation depths (see Equation (\ref{eq_Eax_ay})), the same phase distribution can be observed in the far-zone field \cite{amini_2021}.
In order to have a circularly polarized beam with right-hand polarization state, the coefficients $M_{0x}$ and $M_{0y}$ must have the following relationship: 
\begin{equation}
  M_{0y} = -j M_{0x}
\end{equation}
Fig.\ref{fig:Fig_Ex_Plus3} shows the far-zone pattern, aperture field distribution and mode purity of the multi mode MMLWAs for the case of  $c_{-3} = c_{3} = 0.5$ and $M_{0x} = 0.32$. 
The value of $M_{0x}$ is selected in a way that the synthesized impedance distribution can be easily implemented by common anisotropic unit cells.
The mode purity is measured at $\theta = 8^\circ$, where the antenna gain is maximum. 
Observe that an equal mode distribution is obtained for equal values of $c_{-3}$ and $c_{3}$.
The reason can be explained by the fact that for $l = -3$, $l = +3$, the shape of  far-zone gain are exactly the same and both are maximized at the same spatial angle, except that their vorticities are in opposite directions.  
The far-zone pattern has a petal-like shape with six lobes distributed in the azimuth axis.
In Fig.\ref{fig:Fig_Ex_Plus4} the results for the case of $c_{-4} = c_{4} = 0.33$, $c_{-1} = c_{1} = 0.17$ are plotted. In this case, the angles at which the far-zone amplitudes is maximized are different for topological charges of $l = \pm 4$ and $l = \pm 1$. Observe that at $\theta = 3.6^\circ$ the modes with topological charges of $l = \pm 1$ are dominant and at $\theta = 8.6^\circ$ the modes with $l = \pm 4$ have the maximum purity. ‌Between these two angles, both modes contribute to radiation.
Results indicate that, by tuning the modulation index, the topological charge distribution in the spectral space can be easily controlled.
\section{Antenna implementation}
To implement the metasurface, periodic unit cells may be exploited as constituent elements. They may be sub-wavelength patches printed on a dielectric host medium that support $TM_0$ surface waves. Various geometrical shapes have been proposed in the literature as unit cells \cite{faenzi_2019}-\cite{martini_2014}. The performance of these shapes may be different in terms of anisotropy control, bandwidth, and fabrication tolerance.  In this paper, we have used rectangular patches to meet our tensorial impedance range. 
\begin{figure}
  \centering
       \subfloat[]{%
         \includegraphics[width=0.17\textwidth]{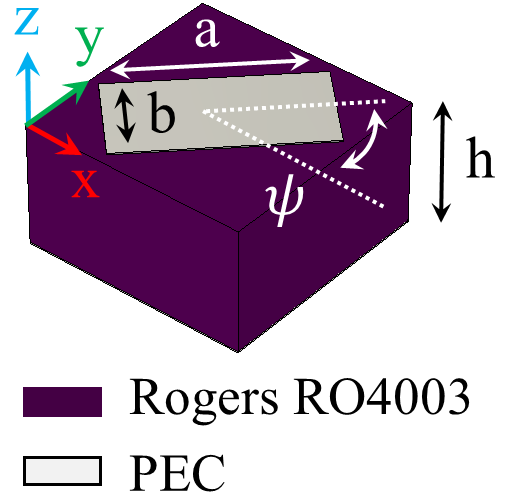}
         \label{fig:Fig_UC}
         }

      \subfloat[]{%
         \includegraphics[width=0.25\textwidth]{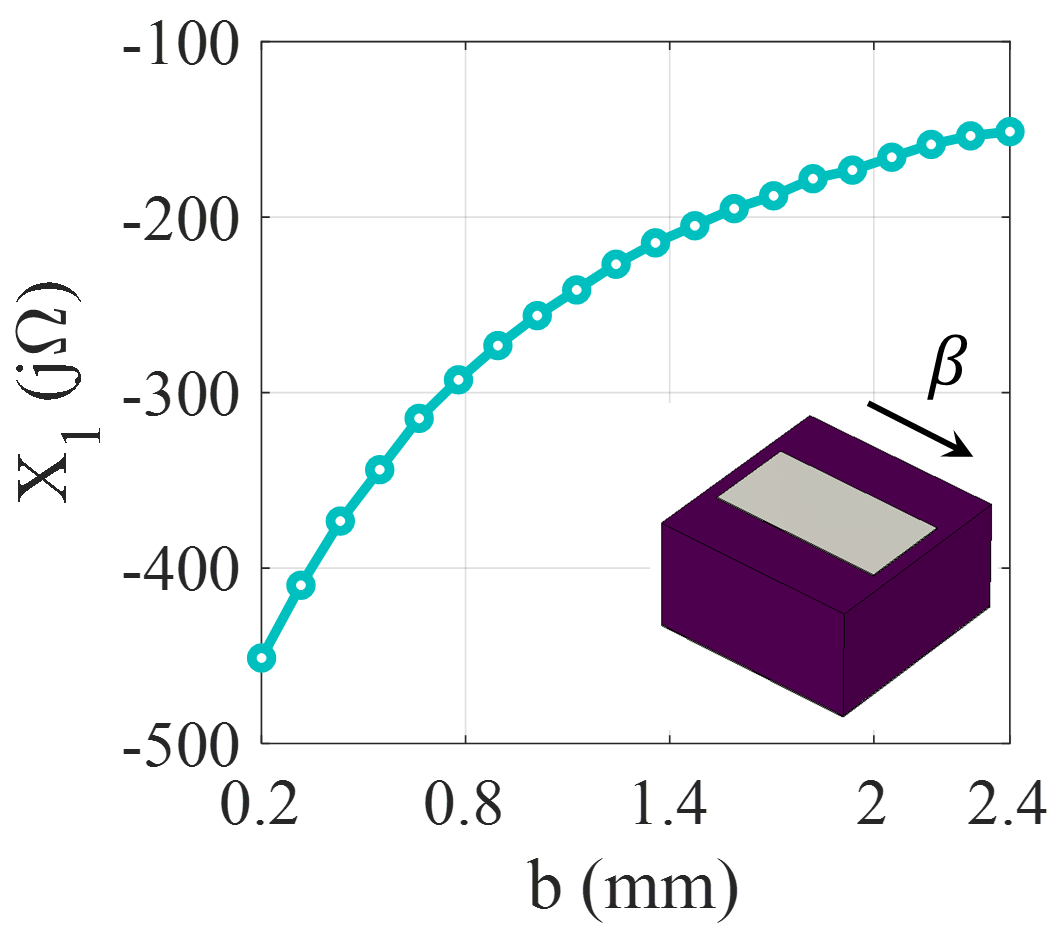}
         \label{fig:Fig_X1}
         }
      \subfloat[]{%
         \includegraphics[width=0.25\textwidth]{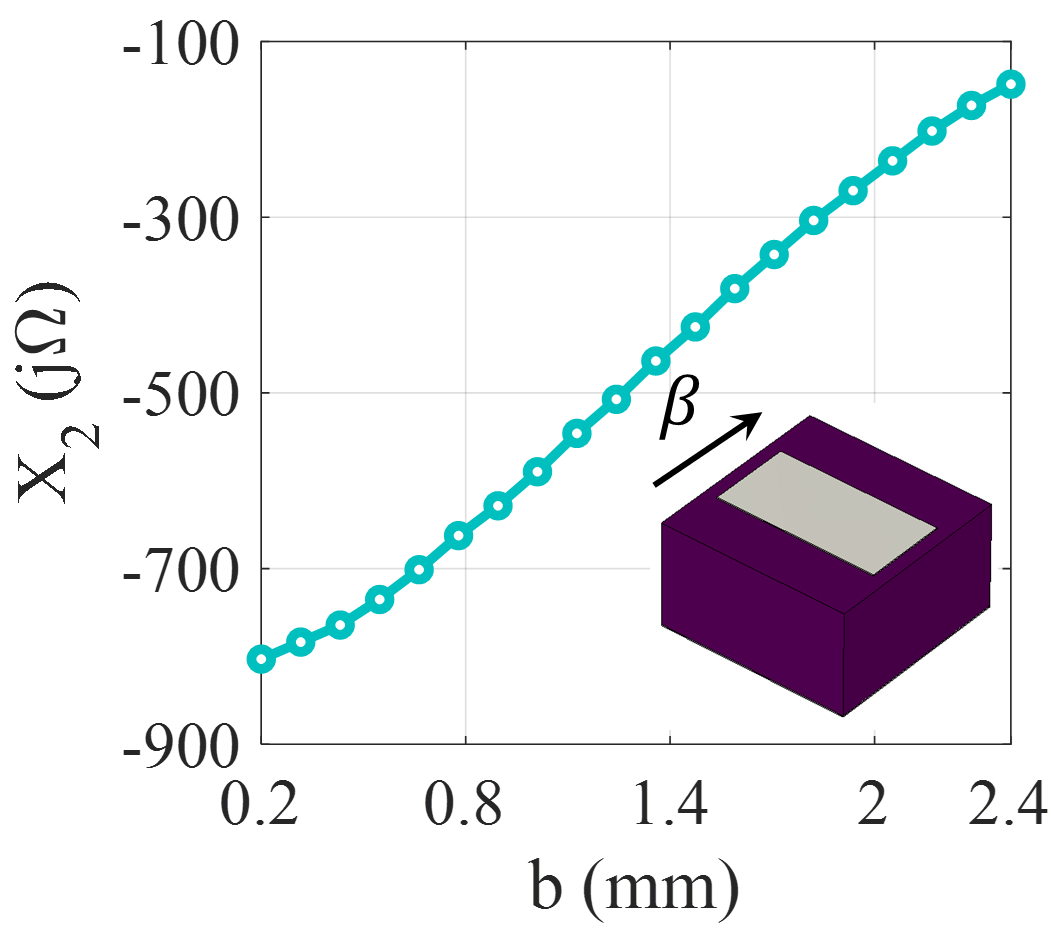}
         \label{fig:Fig_X2}
         }
    \caption{(a) Proposed rectangular patch used for realization of tensorial impedance. (b) and (c) Extracted reactance curves in terms of patch width $b$ for the surface waves propagating along x and y axes, respectively.}
\end{figure}
\begin{figure}
  \centering
      \subfloat[\label{fig:X_rho_rho}]{%
         \includegraphics[width=0.25\textwidth]{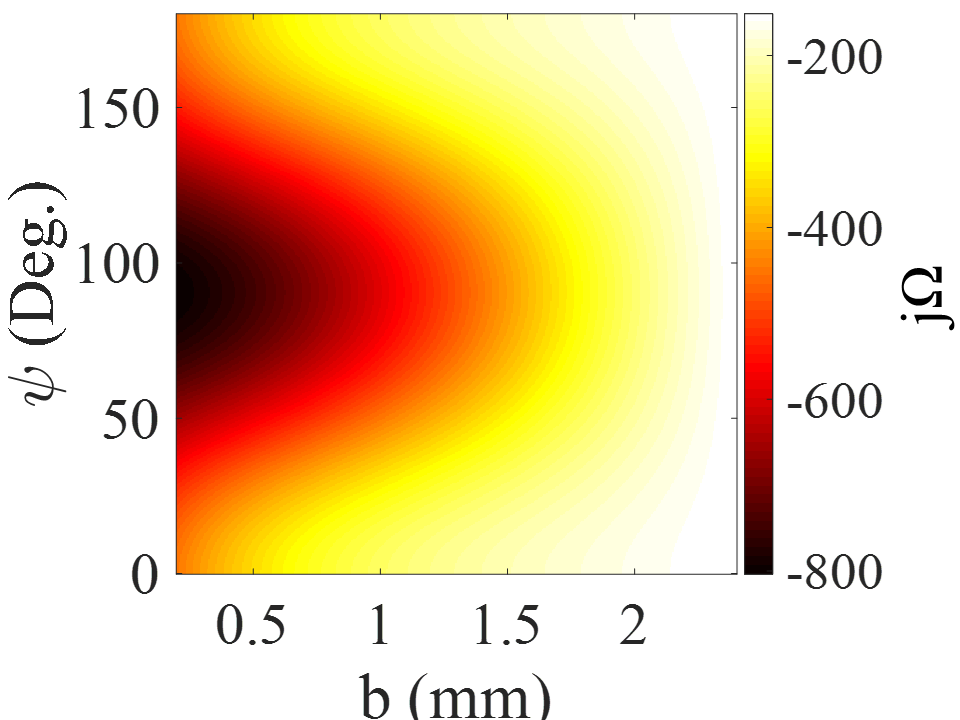}}
      \subfloat[\label{fig:X_rho_phi}]{%
         \includegraphics[width=0.25\textwidth]{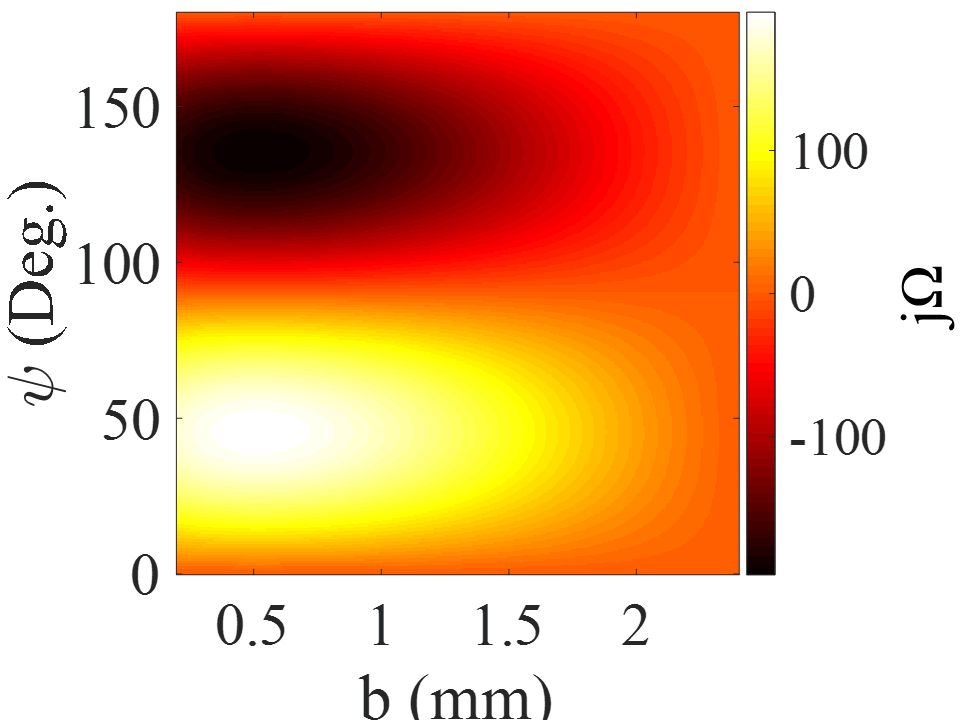}}
    \caption{Reactance maps for the unit cell proposed in Fig.\ref{fig:Fig_UC}. (a) $X_{\rho\rho}$, (b) $X_{\rho\phi}$.}
    \label{fig:XX_rho_rho}
\end{figure}
Fig. \ref{fig:Fig_UC} shows the proposed unit cell.
Rogers RO4003 (with dielectric constant of $\epsilon_r = 3.55$, loss tangent $\tan \delta = 0.0027$ and thickness of $h = 1.524$ mm) is used as the dielectric substrate. In order to effectively model the surface impedance and cover the synthesized impedance range, the period of unit cell is selected as 2.8 mm. The patch length is kept fixed ($a = 2.4$ mm), but its width (namely $b$) and orientation angle (namely $\psi$) are considered to be the variable parameters. For the retrieval of reactance tensor in terms of $b$ and $\psi$, the following equation is used \cite{amini_2021}: 
\begin{equation}
  \underline{\underline{X}}_s = R^T(\psi) \underline{\underline{X}}(b)R(\psi)
\end{equation}
where: 
\begin{equation}
\underline{\underline{X}}(b) = 
\begin{pmatrix}
  X_1(b) & 0\\
  0 & X_2(b)
\end{pmatrix}
\end{equation}
\begin{equation}
  R(\psi) = 
  \begin{pmatrix}
    \cos \psi & -\sin \psi\\
    \sin \psi & \cos \psi
  \end{pmatrix}
\end{equation}
\begin{figure*}
  \centering
       \subfloat[\label{fig:Fig_Test}]{%
         \includegraphics[width=0.5\textwidth]{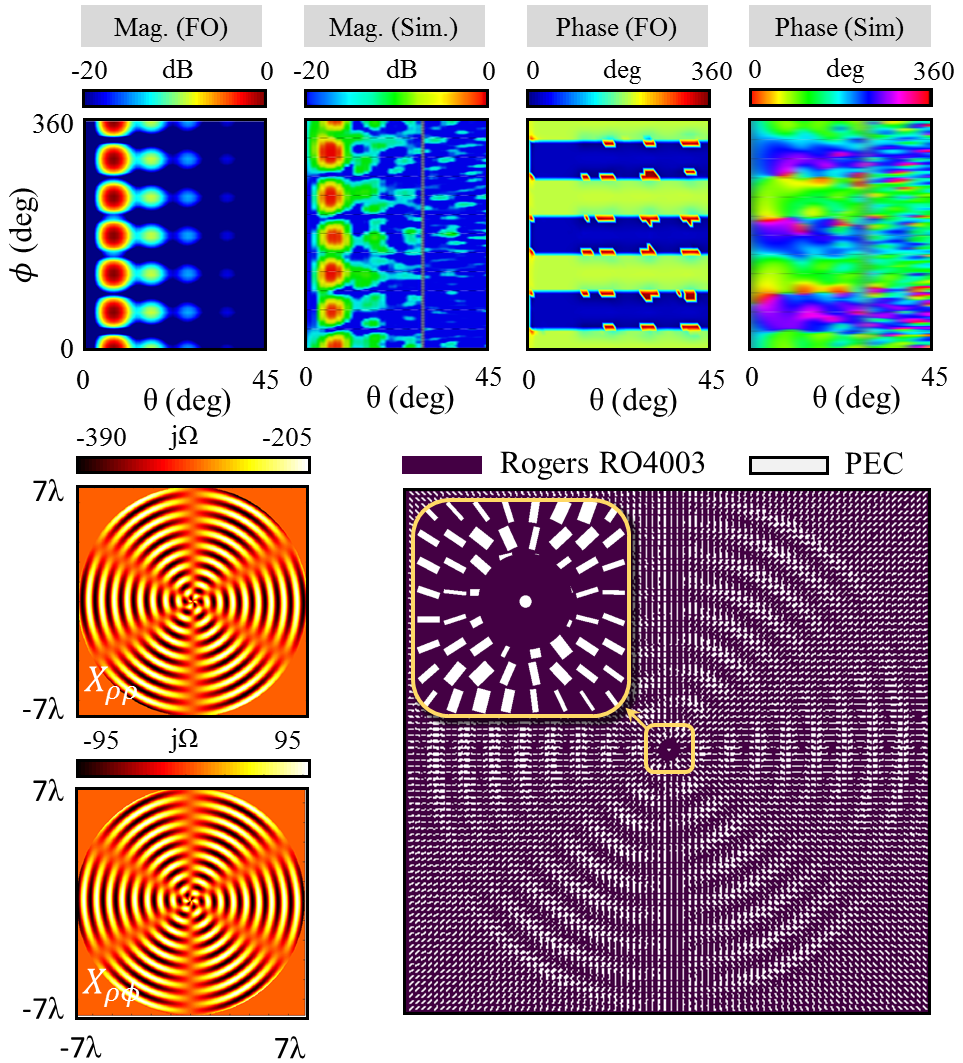}}
      \subfloat[\label{fig:Fig_Test2}]{%
         \includegraphics[width=0.5\textwidth]{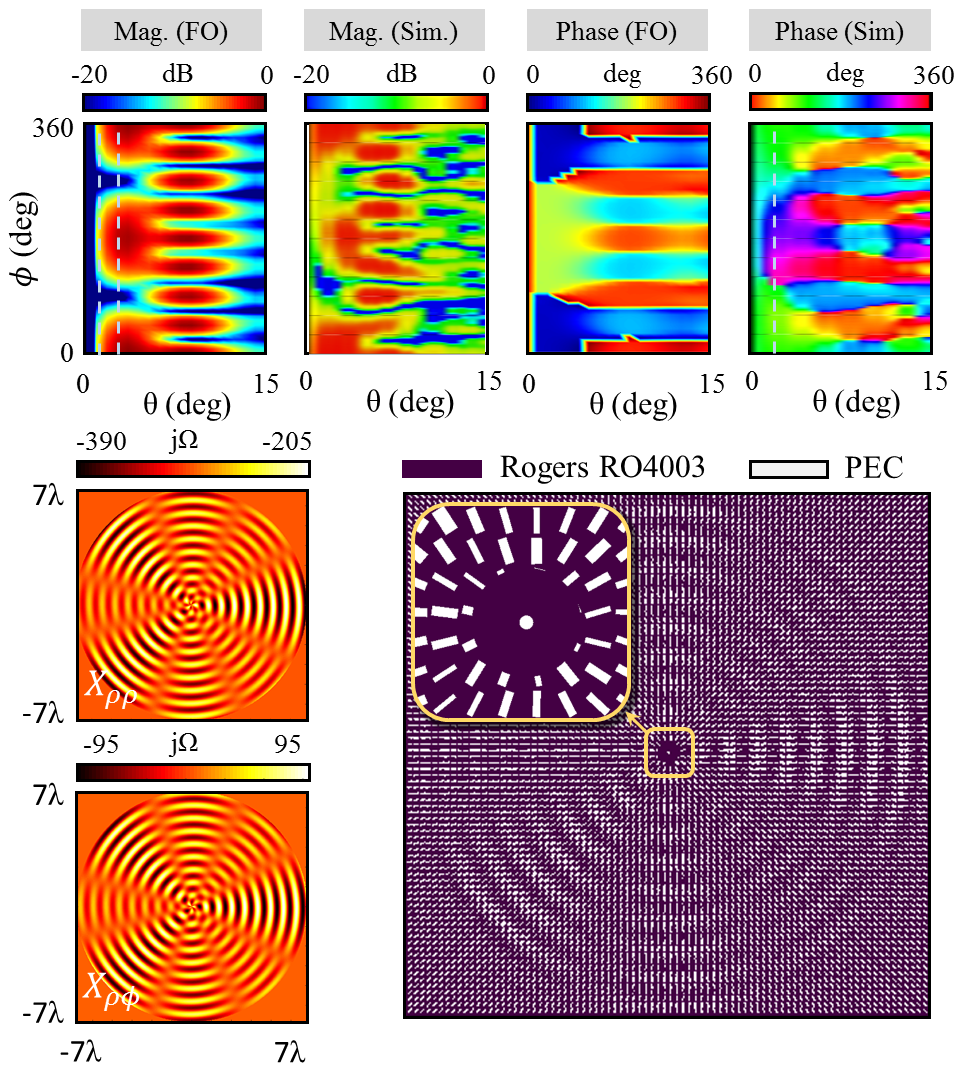}}
    \caption{Simulation and analysis results of far-zone patterns and phase distributions of the proposed MMLWAs for (a) $c_{-3} = c_{3} = 0.5$, (b) $c_{-4}  =  c_{4}= 0.33,  c_{-1} = c_{1} = 0.17$.}
    \label{fig:Fig_Test12}
\end{figure*} 
Note that, $X_1(b)$ and $X_2(b)$ are extracted regard to the wave propagating along the x and y axes, respectively. In both cases, the patch orientation angle is zero. Figs \ref{fig:Fig_X1} and \ref{fig:Fig_X2} show the reactance curves ($X_1$ and $X_2$) by changing the patch width $b$, in which $a$ is equal to $2.4$ mm.
The design maps for $X_{\rho\rho}$ and $X_{\rho\phi}$ versus the patch width and orientation are plotted in Fig. \ref{fig:XX_rho_rho}.
\begin{table*}[]
  \centering
  \caption{Comparison between different schemes for manipulating OAM spectrum.}
  \label{tab:tab1}
  \begin{tabular}{cccccc}
  \hline \hline
  Ref.      & Features                   & Structure (type)  & Feeding type & Analysis method           & Synthesis method                                                                                       \\ \hline
  \cite{yu_2016} & Single-mode & Reflectarray (reflective)     &  Horn ant.           & Full-wave (N. A.)           &  Phase shift distribution                                             \\
  \cite{qin_2018} & Single-mode & Transmitarray (transmissive) & Microstrip ant.   & Full-wave (N. A.) & Phsae shift distribution                                                            \\
  \cite{akram_2019} & Single-mode & Reflectarray (reflective)   &  Horn ant.   & Full-wave (FIT)              &
  Geometric phase concept \\
  \cite{karimipour_2019} & Multi-mode & Reflectarray (reflective)   &  Horn ant.    &  Full-wave (FIT)        &
  Holographic technique \\
  \cite{yang_2021}  & Multi-mode & Reflectarray (reflective)     & Spiral ant.      & Full-wave (FEM)           & Envelope-modulated scheme                                                                              \\
  \cite{yang_2021_tap}  & Multi-mode & Reflectarray  (reflective)    & Spiral ant.      & Huygens-Fresnel principle & \begin{tabular}[c]{@{}c@{}}Envelope-modulated scheme +\\ Multi-mode phase profile\end{tabular}         \\
  \cite{liu_2018}  & Multi-mode & Phase plate (transmissive)    & N. A.       & Fourier series expansion  & Quadratic azimuthal phase variation                                                                    \\
  \cite{zhang_2019}  & Multi-mode  & Plasmonic MS (transmissive)  & Lens  & Full-wave (FDTD)          & \begin{tabular}[c]{@{}c@{}}Combination of geometric phase and \\ Dynamic phase variations\end{tabular} \\
  \cite{cheng_2014}  & Multi-mode & SPP (Transmissive)   &  Waveguide  & Full-wave (FIT)           & Multi-part metasurfaces                                                                                \\
  This work & Multi-mode & MMLWA (leaky-wave)   &  Monopole & Flat Optics               & Aperture field estimation method                                                                       \\ \hline
  \end{tabular}
  \end{table*}
\section{Simulation results}
To implement the multi-mode OAM generators in  section \ref{subsec:multi_mode}, the unit cell proposed in Fig. \ref{fig:Fig_UC} is used as the constituent pixel. The modulated metasurface is excited by a vertical monopole at the origin. Fig.\ref{fig:Fig_Test12} shows the surface impedance distributions, realized MMLWAs, and their corresponding simulation results. 
The simulation is performed by the  transient solver in CST Microwave Studio software.
The average reactance ($X_0$) and modulation depth ($M_{0x}$) are selected as $0.8\eta_0$ and $0.32$, respectively in order to satisfy the reactance variations in Fig.\ref{fig:XX_rho_rho}.
The variation range of $X_{\rho\rho}$ is -305 to -205 $j\Omega$ and the reactance $X_{\rho\phi}$ varies from -95 to +95 $j\Omega$. 
The overall size of metasurfaces is selected as $14\lambda \times 14\lambda$ which is large enough so that the surface waves may effectively be converted into  space waves carrying OAM modes.
Note that decreasing the modulation depth ($M_{0x}$), reduces the leakage factor \cite{patel_2011} and consequently increases the beam convergence. This parameter can be used as a degree of freedom to control the beam convergence of MMLWA. 
For the case of $c_{-3} = c_{3} = 0.5$ (Fig.\ref{fig:Fig_Test}), the petal-like beam with six lobes is observed which confirms the coaxial superposition of $l = -3$ and $l = +3$ modes.
The simulation and analytical results of patterns for the case of $c_{-4} = c_4 = 0.33$ and $c_{-1} = c_1 = 0.17$ are compared in Fig. \ref{fig:Fig_Test2}. The beam has two lobes at $\theta = 3.6^\circ$ and eight lobes at $\theta = 8.6^\circ$ indicating that at lower angles the topological charges of $l = \pm 1$ are dominant and at greater angles the topological charges of $l = \pm 4$ have the maximum energies.

Table \ref{tab:tab1} compares  different methods of synthesis and analysis of OAM wave generators. 
references \cite{yu_2016}-\cite{akram_2019} are designed for single-mode operation, while other references in Table \ref{tab:tab1} have multi-mode properties.
Observe that  all references have used reflective or transmissive configurations. The main limitation of these structures is their requirements for protruding feed systems, which greatly increases the overall dimensions of the antenna and makes them unsuitable for integrated and low-profile transceivers.
Furthermore, dielectric losses in multi-layer transmitarrays may be a challenging issue for the implementation of low loss antennas.
In our work, we have used the modulated leaky-wave antennas to achieve a multi-mode vorticity.  Leaky-wave antennas, in addition to having the advantages of reflectarrays/transmitarrays, use embedded feeding systems that significantly enhance the compactness of structure. 
In all references except  \cite{yang_2021_tap} and \cite{liu_2018}, the full-wave methods have been used for the analysis of entire structure. Full-wave methods have some limitations in simulating large-scale structures such as high memory usage and CPU time consumption , which make them impractical to analyze with conventional computers. Using the analytical framework, in addition to low memory consumption, gives the designer deep insight into the radiator performance.

\section{Conclusion}
In this work, a theoretical method for designing leaky-wave metasurface antennas with arbitrary combination of OAM modes is presented. In this method, by precisely controlling the modulation index, the desired distribution of mode energies in the spectral space can be achieved. This technique can be applied for large-scale structures where full-wave simulators may have some challenges in the simulation. Owing to the fact that the designed antenna is a leaky-wave type, it can be well integrated with microelectronic devices.

\ifCLASSOPTIONcaptionsoff
  \newpage
\fi

%






\end{document}